\documentclass[review,number,sort&compress]{elsarticle}

\usepackage{hyperref}
\usepackage{csquotes}
\usepackage[capposition=bottom]{floatrow}
\usepackage{amsmath}
\usepackage[]{natbib}
\usepackage{subfigure}
\usepackage{lineno}
\usepackage{csquotes}
\usepackage{graphicx}
\usepackage{caption}
\usepackage{afterpage}
\usepackage{placeins}

\begin{document}

\begin{frontmatter}

\title{Neutron spectroscopy with the Spherical Proportional Counter based on nitrogen gas}

\author[a]{E. Bougamont}
\author[a,d]{A. Dastgheibi}
\author[a]{ J. Derre}
\author[a]{J. Galan}
\author[b]{G. Gerbier}
\author[a]{I. Giomataris}
\author[a]{M. Gros}

\author[a,c]{I. Katsioulas\corref{mycorrespondingauthor}}
\cortext[mycorrespondingauthor]{Corresponding author}
\ead{ioannis.katsioulas@cea.fr}

\author[a]{D. Jourde}
\author[a]{P. Magnier}
\author[a]{ X.F. Navick}
\author[a]{T. Papaevangelou}
\author[c]{I. Savvidis}
\author[a]{G. Tsiledakis}

\address[a]{IRFU, Centre d'\'etudes de Saclay, Gif sur Yvette, France}
\address[b]{Queen’s University, Kingston, Canada}
\address[c]{Aristotle University of Thessaloniki, Thessaloniki, Greece}
\address[d]{LSM, CNRS/IN2P3, Universit\'e Grenoble-Alpes, Modane, France}


\begin{abstract}
A novel large volume spherical proportional counter, recently developed, is used for neutron measurements. The pure $N_{2}$ gas is  studied for thermal and fast neutron detection, providing a new way for neutron spectroscopy. The neutrons are detected via the ${}^{14}N(n,p){C}^{14}$ and ${}^{14}N(n,\alpha){B}^{11}$ reactions. The detector is tested for thermal and fast neutrons detection with ${}^{252}Cf$ and ${}^{241}Am-{}^{9}Be$ neutron sources. The atmospheric neutrons are successfully measured from thermal up to several MeV, well separated from the cosmic ray background. A comparison of the spherical proportional counter with the current available neutron counters is also presented. 
\end{abstract}

\end{frontmatter}

\section{Introduction}

\begin{figure}[h]

  \centering
  \subfigure[]{\includegraphics[width=0.46\textwidth]{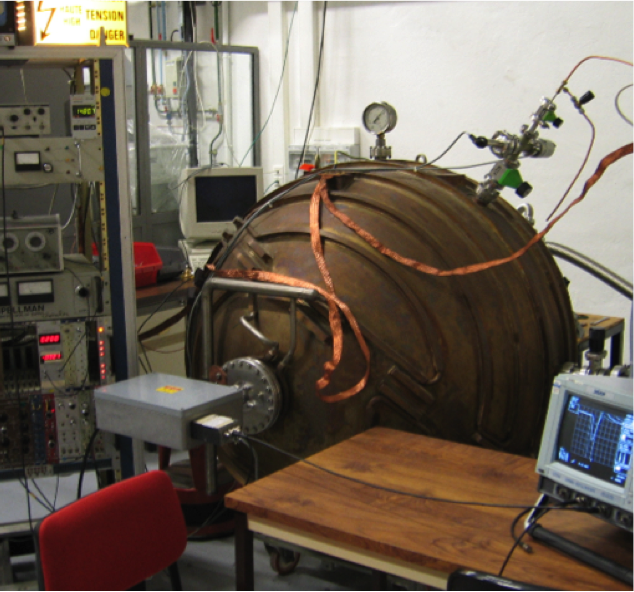}}
  \quad
  \subfigure[]{\includegraphics[width=0.46\textwidth]{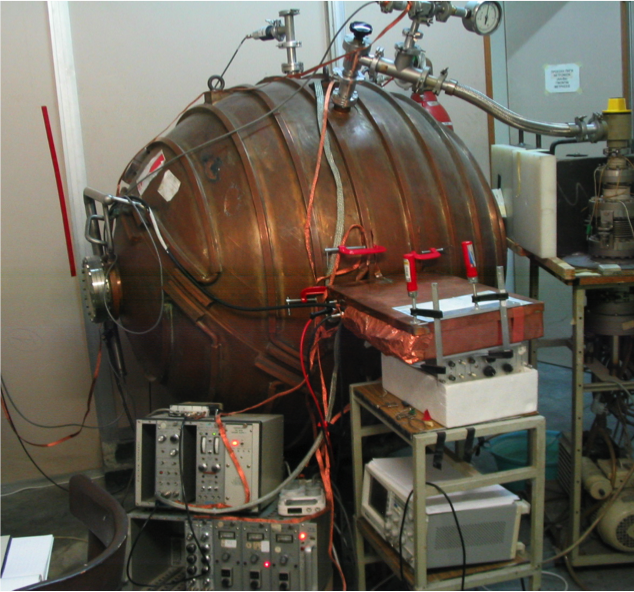}}
  \caption{Pictures of the SPC and the electronics setup in (a) the CEA, Saclay and (b) the Aristotle University of Thessaloniki.}
  \label{spc}
\end{figure}

The Spherical Proportional Counter (SPC) (figure \ref{spc}) is a novel concept with very promising features, among which is the possibility of easily instrumenting large target masses with good energy resolution and low energy threshold.  The natural radial focusing of the spherical geometry allows collecting and amplifying the deposited charges by a simple and robust detector using a single electronic channel to read out large gaseous volumes \citep{giomataris1,giomataris2,aune}. The sensitivity of the detector is obtained using the Rn daughters decay to alpha particles in the energy range of $5.3-7.7\ MeV$. The low energy calibration of the detector from $100\ eV$ up to several tenths of keV is achieved by using X-rays and a UV lamp \citep{giomataris3,andriamonje}.  
The detector has been successfully used as a neutron detector based on the ${}^{3}He(n,p)H^{3}$ reaction \citep{savvidis}. At the present work the SPC has been optimized for operation with pure nitrogen ($N_{2}$), which presents an interesting neutron converter, permitting neutron spectroscopy up to  $20\ MeV$. The thermal and fast neutrons are detected via the ${}^{14}N(n,p){C}^{14}$ reaction and fast neutrons with energy  greater than 1.7 MeV via the${}^{14}N(n,\alpha){B}^{11}$ reaction. The low background of the detector and the possibility to separate the $\gamma$ ray pulses from proton and alpha particle pulses (through pulse shape analysis) increases the sensitivity for neutron detection. This detector can successfully measure very low thermal neutron fluxes, enabling measurements of neutron energy spectra up to several MeV at ground and underground level \citep{savvidis}.

\section{The detector}

\begin{figure}[!h]
\centering
\includegraphics[width=90mm]{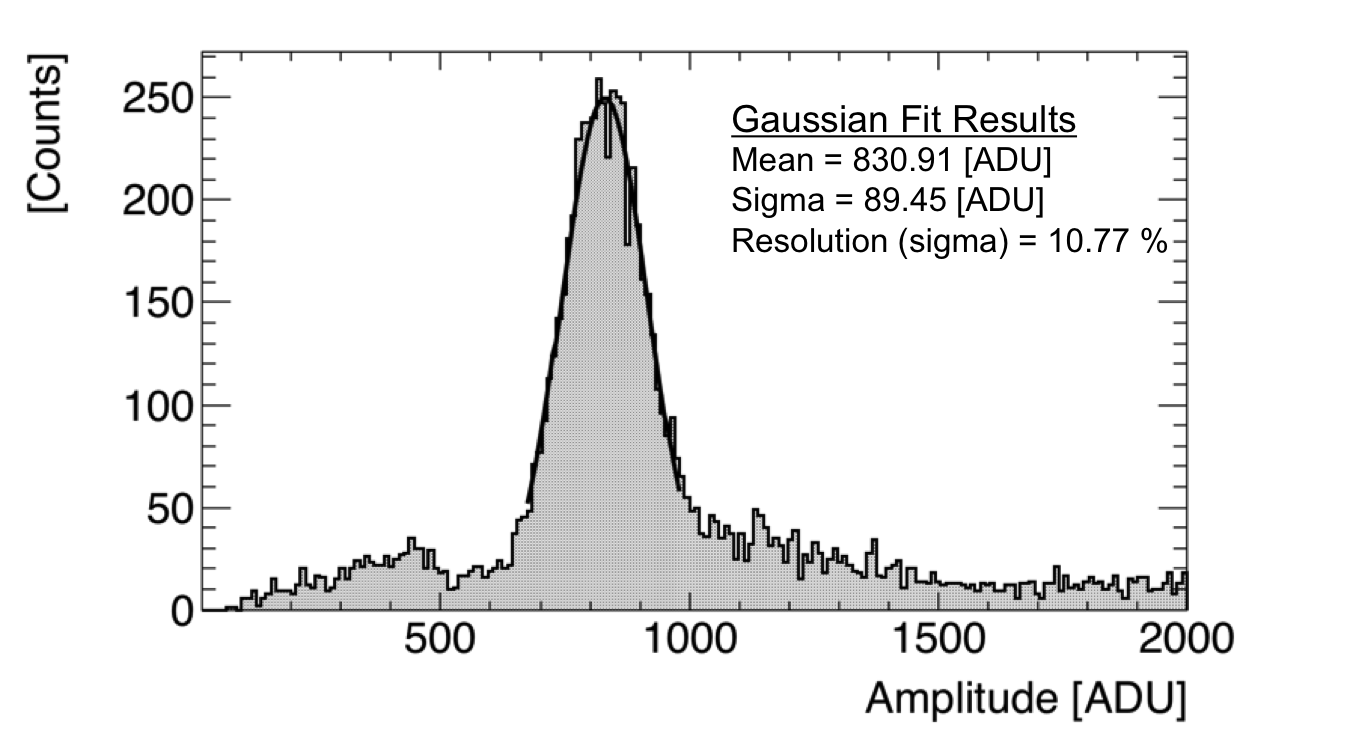}
\caption{Spectrum of the ${}^{55}Fe$ source obtained with the spherical detector filled with nitrogen ($N_{2}$) gas at $500\ mbar$.}
\label{FeFit}
\end{figure}

The detector consists of a large spherical copper vessel, $1.3\ m$ in diameter and $6\ mm$ thick. It is operated in a sealed mode, after being well pumped (up to $10^{-8}\ mbar$) and filled with a gas mixture at pressure of several hundred $mbar$ up to 5 bar. An outgassing in the order of $10^{-9} \ mbar/s$ is necessary for the amplification stability, because the presence of the $O_2$ in the drift volume induces electron attachment. A small stainless ball (sensor), usually of $14\ mm$ in diameter, fixed in the center of the spherical vessel by a stainless steel rod, acts as an electrode with positive high voltage (anode) in respect to the grounded vessel (cathode) and as a proportional amplification counter. Details about the design and principle of the SPC can be found at \citep{giomataris3} where a complete description of the detector exists. 
In order to fulfill the requirements for operation with pure $N_{2}$ the $14-mm$ ball had to be replaced by a smaller $8-mm$ or $3-mm$ ball. In addition, the metallic ball had been replaced by a silicon one to reduce the negative effects of occasional sparks. These improvements were necessary to achieve proportional amplification in pure $N_{2}$, a process occurring at extremely high electric fields. Figure \ref{FeFit} shows a typical spectrum produced by a ${}^{55}Fe$ source at $500\ mbar$ $N_{2}$ pressure, the obtained resolution ($\sigma$) is reasonably good at about $11\%$. 

\section{The neutron detection}

\begin{figure}[!h]
\centering
\includegraphics[width=90mm]{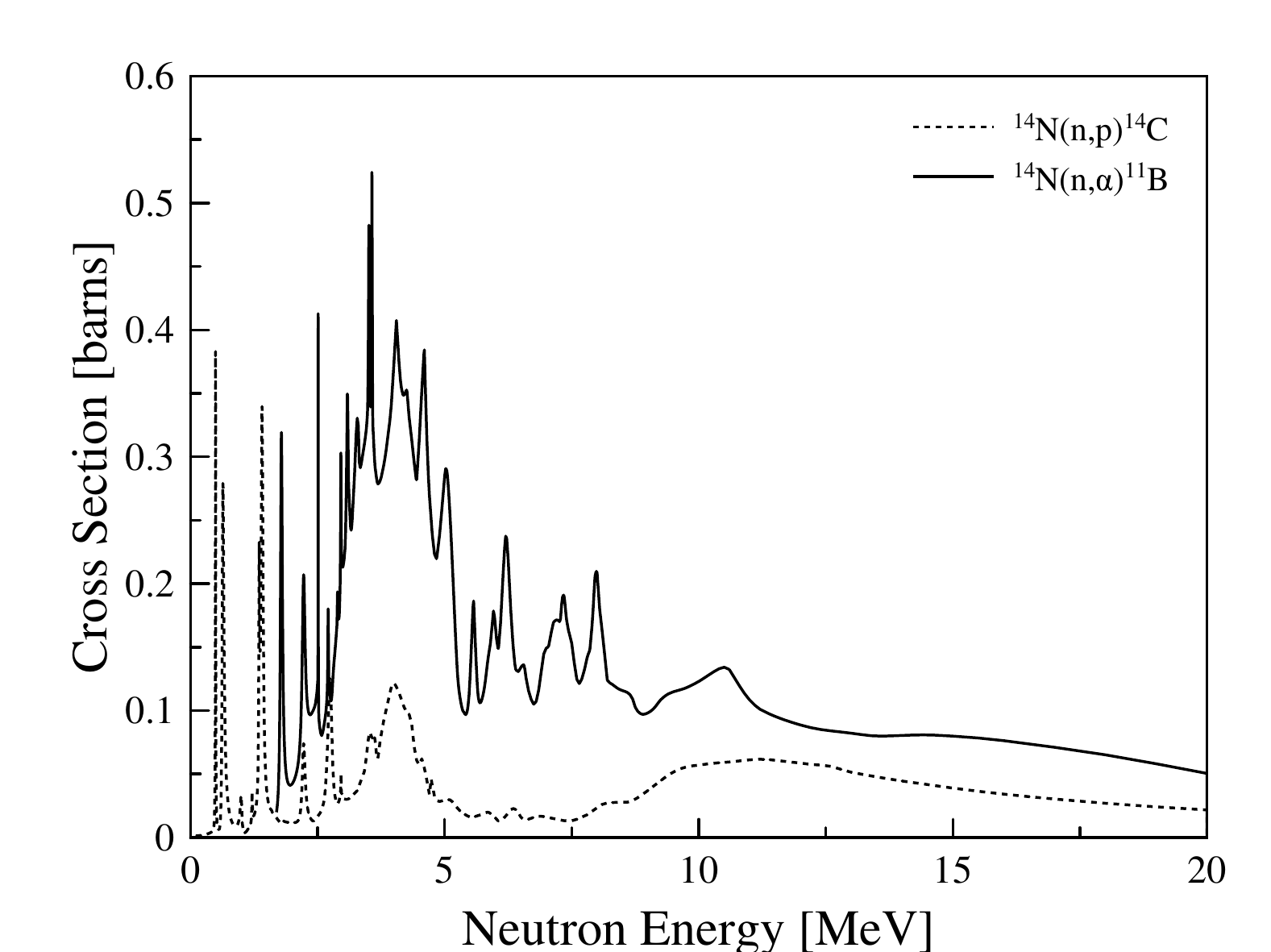}
\caption{The cross sections of the reactions ${}^{14}N(n,\alpha){B}^{11}$ and  ${}^{14}N(n,p){C}^{14}$ for neutrons of energy up to $20\ MeV$ \citep{endfvII}.}
\label{cs1}
\end{figure}

The detector is filled with pure $N_{2}$ at pressure up to $500\ mbar$. Thermal neutrons interact with the ${}^{14}N$ nucleus via the reaction
\begin{equation*}
{}^{14}N + n \rightarrow {}^{14}C + p + 625.87 keV                 
\end{equation*}
The energy of the exothermic reaction is shared as kinetic energy between the ${}^{14}C$ and the proton with $E_C = 41.72\ keV$ and $E_p = 584.15\ keV$. Fast neutrons can be detected via the 
\begin{equation*}
{}^{14}N + n \rightarrow {}^{11}B + \alpha - 158 keV             
\end{equation*}
reaction up to $E_n = 20\ MeV$. The cross section for the thermal neutron detection is $1.83\ barn$ \cite{gledenov} and for fast neutron detection is presented in figure \ref{cs1}. The endothermic reaction is the one with the most significant contribution to fast neutron detection for incident neutron energies greater than $1.7\ MeV$, since the cross section of the ${}^{14}N(n,\alpha){B}^{11}$ reaction is higher than that of the ${}^{14}N(n,p){C}^{14}$ reaction after this energy value. 

\begin{figure}[h]
  \centering
  \subfigure[]{\includegraphics[width=90mm]{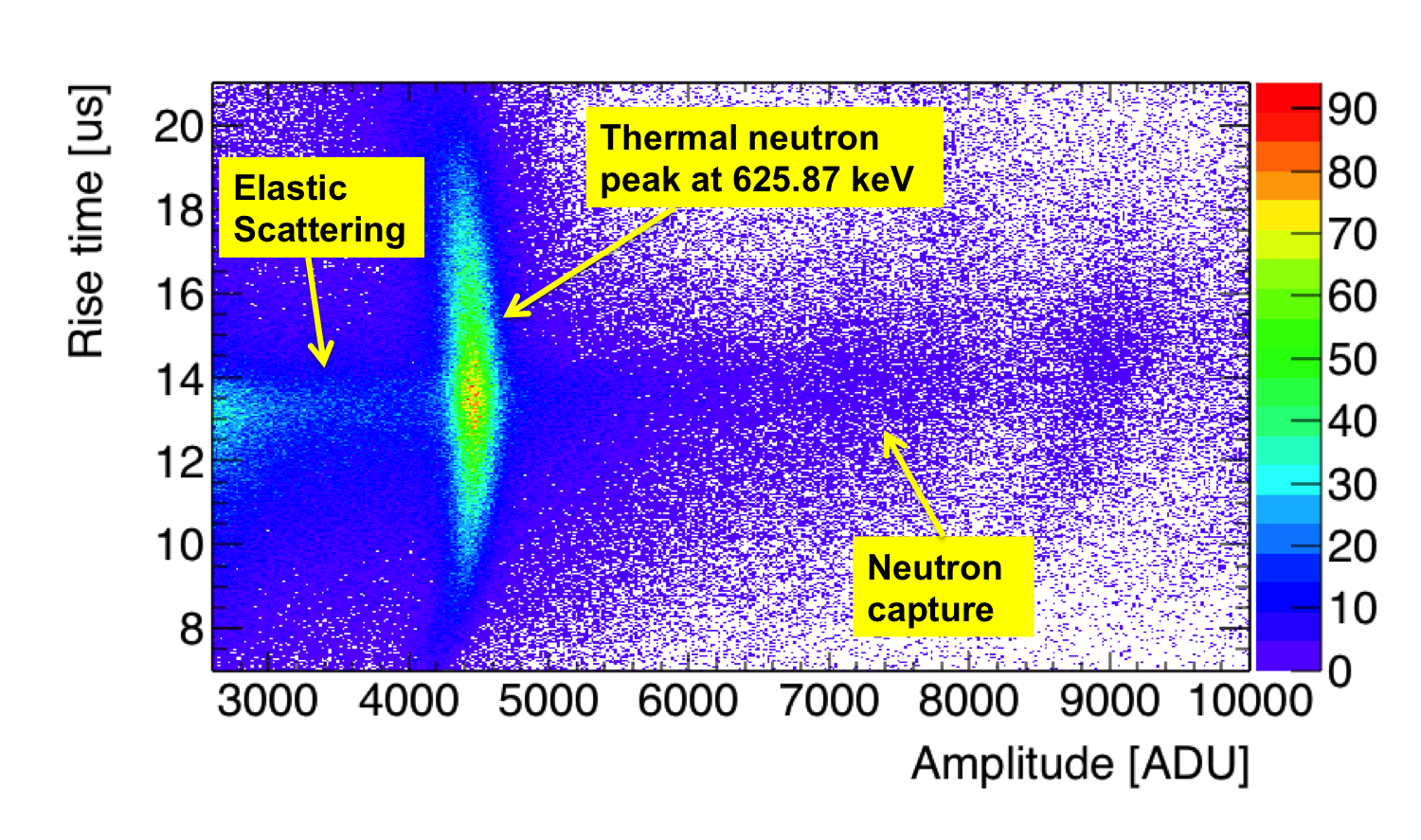}\label{thermala}}
  \\
  \subfigure[]{\includegraphics[width=90mm]{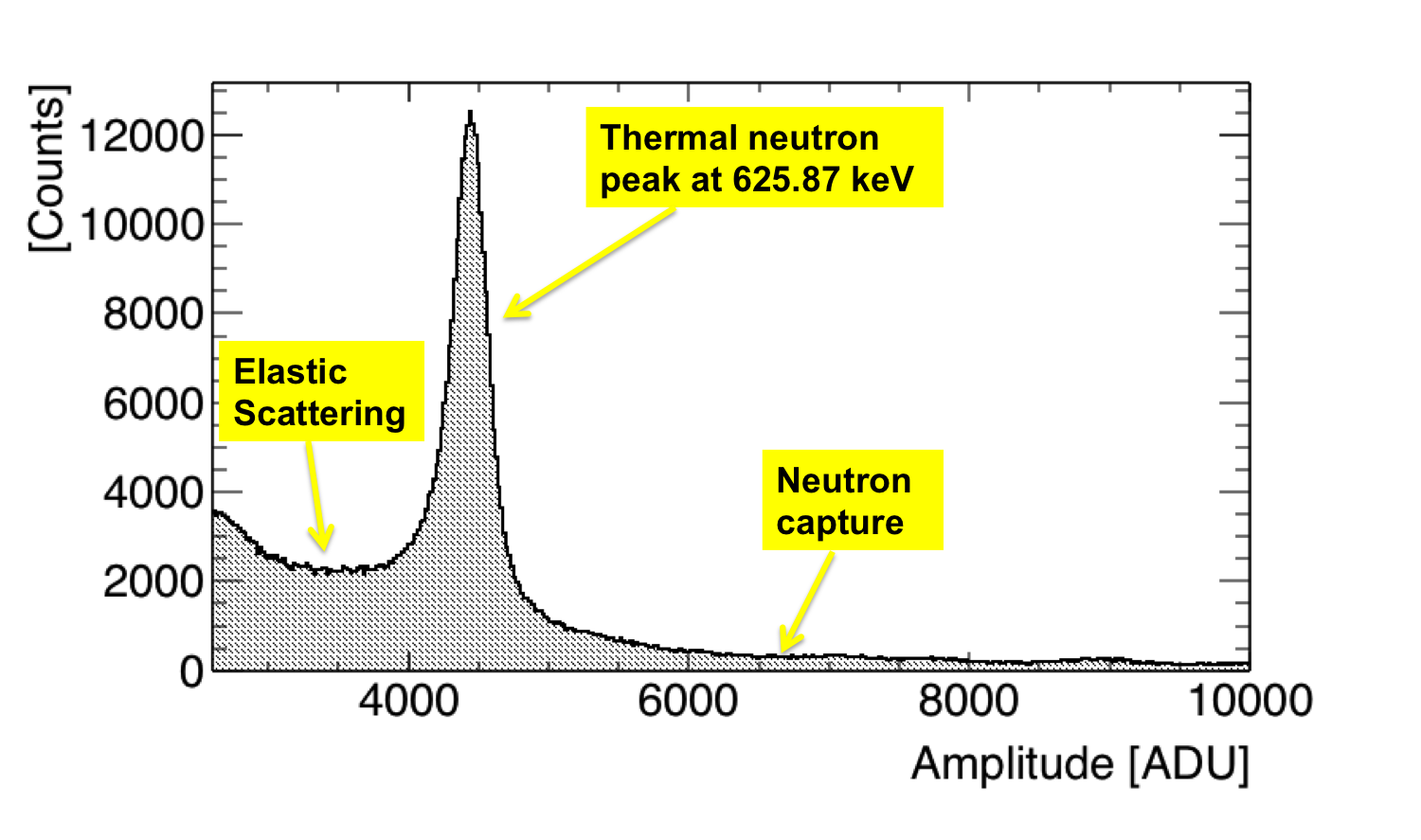}\label{thermalb}}
  \caption{Thermal neutron measurements after irradiation by a ${}^{252}Cf$ source. The gas used was $400\ mbar$ of pure $N_{2}$, the sensor was $3-mm$ in diameter, the high voltage was set at $6000\ V$ providing a gain of $100$. The measurements presented are (a) the rise time as a function of the amplitude, (b) the amplitude spectrum in ADC units. The detection energy threshold was $370\ keV$. }
  \label{thermal}
\end{figure}

The neutron detection results are presented in figure \ref{thermala} as the rise time versus amplitude of the pulse and the amplitude spectrum in ADC units (figure \ref{thermalb}). The rise time of the pulse provides the depth of the ionization electrons produced in the gas and the amplitude the deposited energy. The gas used was pure $N_{2}$ at $400\ mbar$ pressure. The detector was irradiated by a ${}^{252}{Cf}$ source, cased inside a polyethylene (PE) half box (that covered the back solid angle), placed $30\ cm$ away from the surface of the spherical detector. This technique was used to thermalize a part of the fast neutron spectrum of the ${}^{252}{Cf}$ source. The peak at $625.87\ keV$ belongs to thermal neutrons detected through the ${}^{14}N(n,p){C}^{14}$ reaction and is well separated from the cosmic rays and the nuclear recoils.  The fast neutrons are detected through both the $(n,p)$ and the $(n,\alpha)$ reactions and appear to the right of the thermal peak. The recoil nuclei, which are produced by the fast neutron elastic scattering, appear in the region left (below $~4000\ ADU$) from the thermal peak at lower energies.

\begin{figure}[h]
  \centering
  \subfigure[]{\includegraphics[width=90mm]{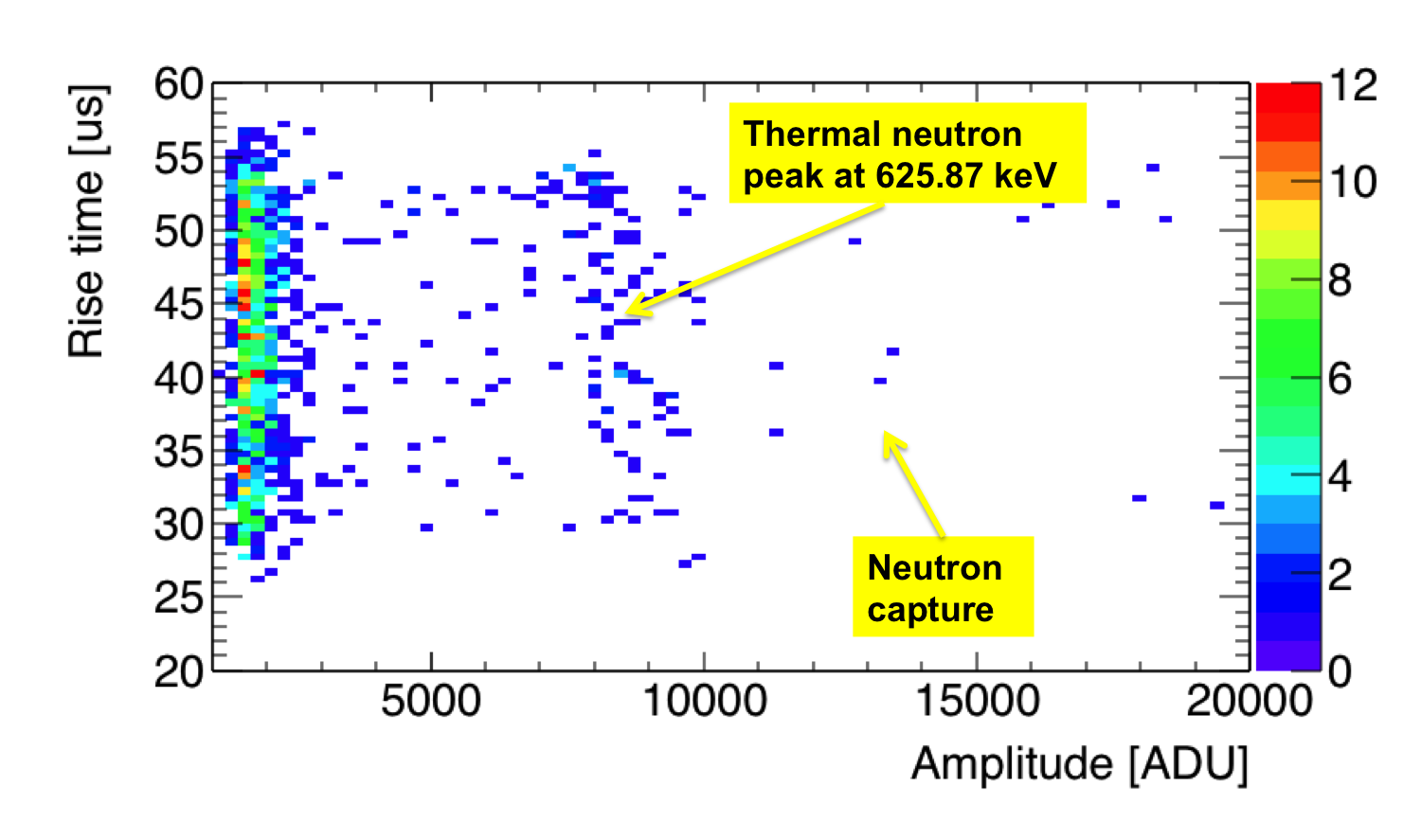}\label{thermal2a}}
  \\
  \subfigure[]{\includegraphics[width=90mm]{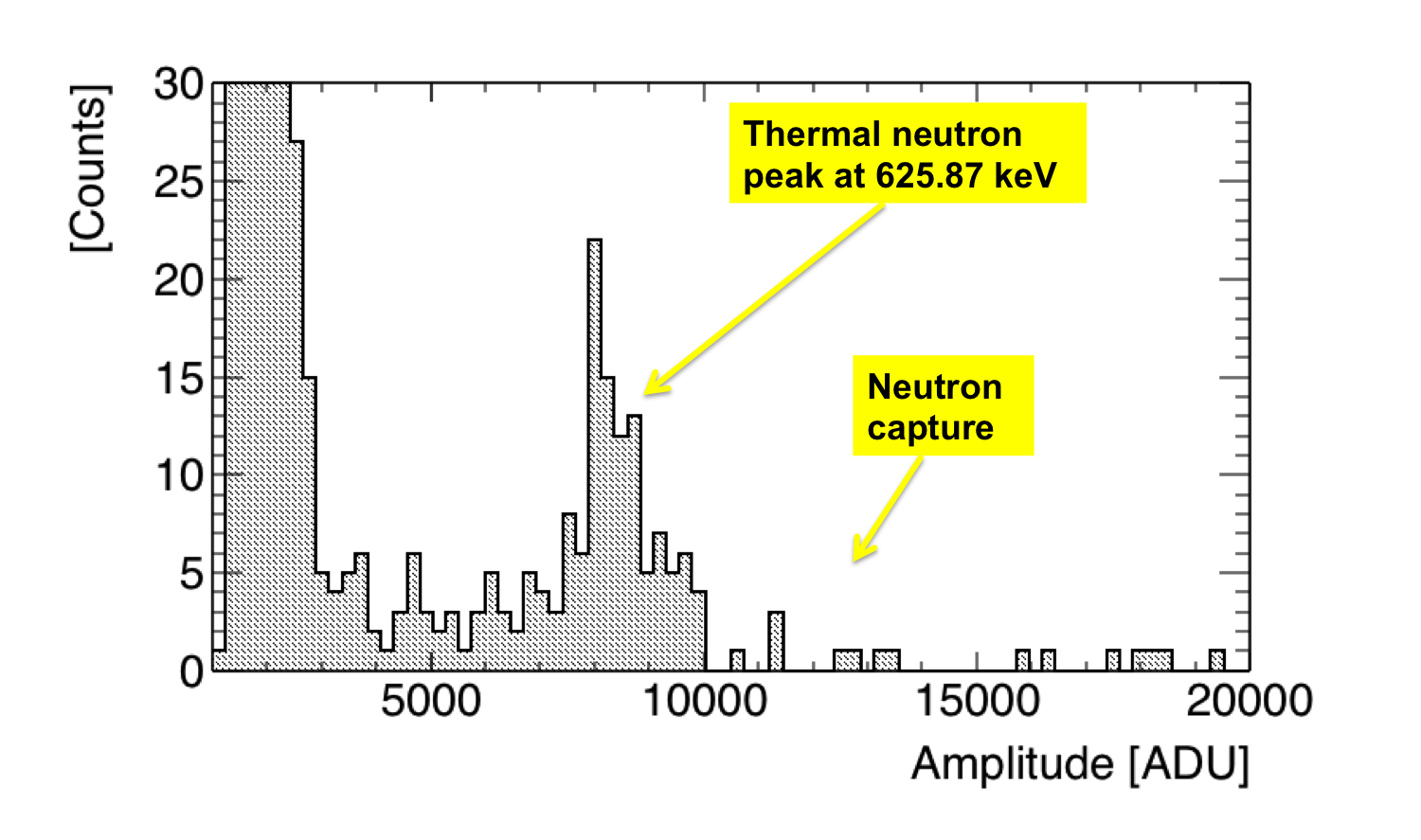}\label{thermal2b}}
  \caption{Atmospheric neutron detection with $500\ mbar$ pure $N_2$ gas, with a sensor of $3\ mm$ in diameter and a high voltage of $6200\ V$. The measurements presented are (a) the rise time as a function of the amplitude (b) the amplitude spectrum in ADC units. The detection energy threshold was 75 keV and the rate in the $0.9\ MeV$ $-$ $2.2\ MeV$ region was $2.8\ mHz$. This rate is considered to be the total background rate (considering muon and neutrons incduced events) for fast neutron measurements.}
  \label{thermal2}
\end{figure}

The atmospheric neutrons have been also successfully measured with pure $N_2$ at $500\ mbar$. The thermal neutron peak and the fast neutrons measured are presented in figure \ref{thermal2} after 16 hours of acquisition.

\FloatBarrier
\section{Fast neutrons measured with ${}^{252}Cf$ and ${}^{241}Am-{}^{9}Be$ neutron sources and the wall effect}

\begin{figure}[h]
  \centering
  \subfigure[]{\includegraphics[width=110mm]{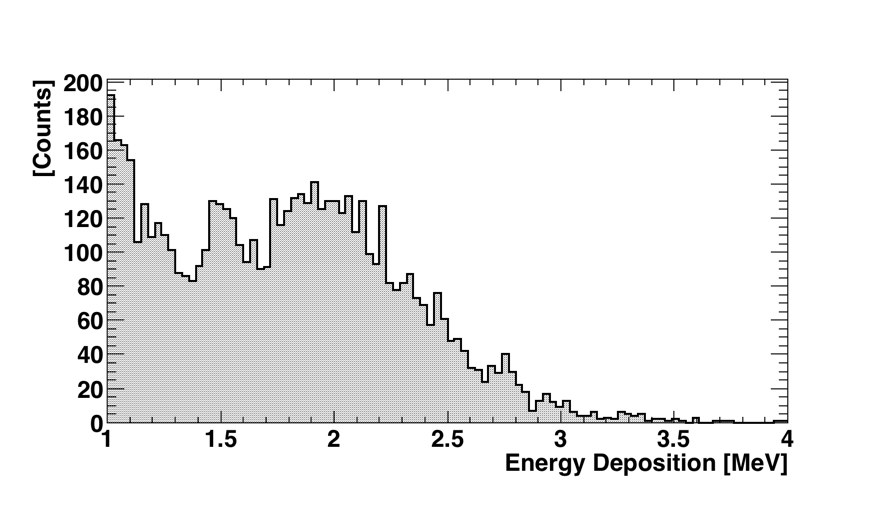}\label{cfa}}
  \\
  \subfigure[]{\includegraphics[width=90mm]{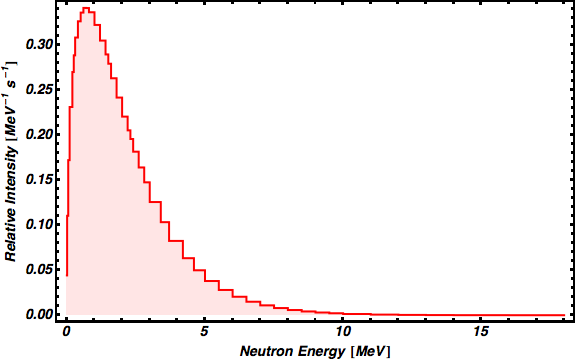}\label{cfb}}
  \caption{(a) Measured energy spectrum after irradiation by fast neutrons from a ${}^{252}Cf$ neutron source ($3-mm$ sensor ball, $N_2$ gas filling at $400\ mbar$). The $(n,p)$ cross section coincidence peaks appear in the measured energy deposition spectrum. (b) ${}^{252}Cf$ source neutron energy spectrum \citep{iso}. Taking into account the activity of the source and its position relative to the SPC, the efficiency was estimated to be $\sim 10^{-3}$.}
  \label{cf}
\end{figure}

Long irradiation time measurements with ${}^{252}Cf$ ($ 6.4 \cdot 10^{4} \ Bq$) and ${}^{241}Am-{}^{9}Be$ ($6.0 \cdot 10^{4}\ Bq$) neutron sources have been performed, to study fast neutron detection.  In figure \ref{cfa} we present the measured energy distribution of the events produced by the fast neutrons of the ${}^{252}Cf$ source (figure \ref{cfb}. For the measurement, the detector was filled with pure $N_2$ at $400\ mbar$ and a sensor with a silicon ball, $3\ mm$ in diameter, was used. The ${}^{252}Cf$ neutrons have a distribution with a maximum close to $1.4\ MeV$ (figure \ref{cfb}). In the energy range of the ${}^{252}Cf$ neutrons, the $(n,p)$ reaction plays a dominant role (figure \ref{cs1}). In the energy spectrum (figure \ref{cfa}), we see the characteristic $(n,p)$ cross section peaks and the accumulation of signals at lower energies because of the wall effect \citep{wall1}\citep{wall2}. In order to measure the neutron energy, the produced particles from both $(n,p)$ and $(n,\alpha)$ reactions must deposit all their energy in the drift volume. If the reaction takes place close to the vessel wall, it is possible for one of the charged particles or both of them to hit the wall and to lose a part of their energy. This is known as the wall effect and leads to wrong estimation of the neutron energy.

\begin{figure}[h]

  \centering
  \subfigure[]{\includegraphics[width=110mm]{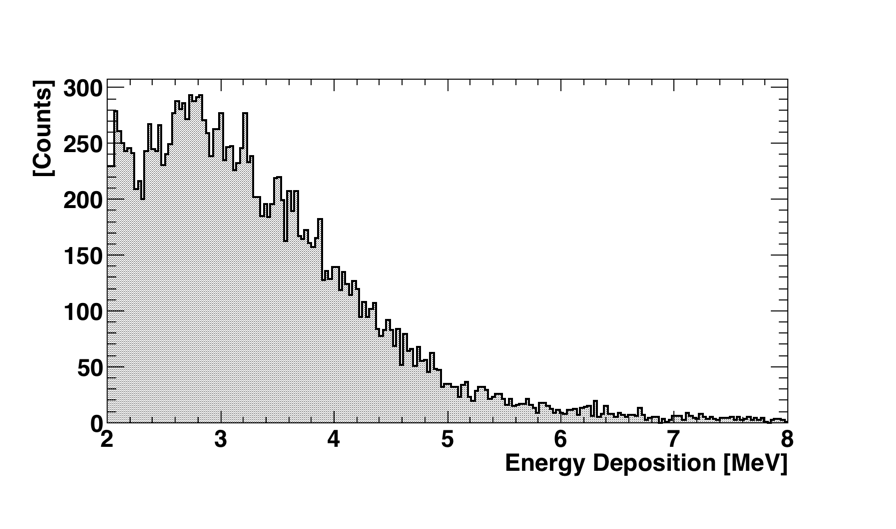}\label{ambea}}
  \\
  \subfigure[]{\includegraphics[width=90mm]{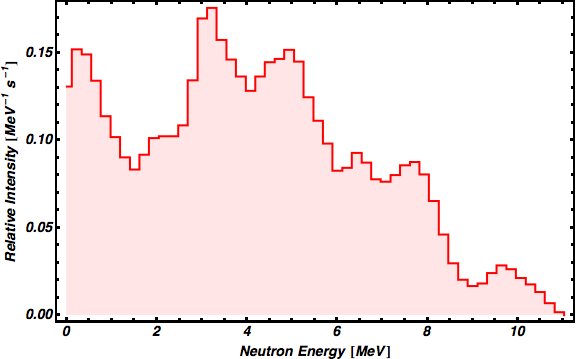}\label{ambeb}}
  \caption{(a) Measured energy spectrum after irradiation with fast neutrons from a ${}^{241}Am-{}^{9}Be$ source ($8-mm$ sensor ball,  $N_2$ gas filling at $200 \ mbar$). There is a translation of the spectrum to lower energies because of the wall effect. (b) ${}^{241}Am-{}^{9}Be$ source neutron energy spectrum \citep{iso}.}
  \label{ambe}
\end{figure}

In figure \ref{ambea} we present the measurements after irradiation by an ${}^{241}Am-{}^{9}Be$ neutron source. The sensor used was an $8-mm$ metallic ball and pure $N_2$ gas at $200\ mbar$ pressure. Since the neutrons of an ${}^{241}Am-{}^{9}Be$ source (figure \ref{ambeb}) have higher energy compared to the neutrons of a ${}^{252}Cf$ source, most events are coming from the $(n,\alpha)$ reaction. As in the previous figure, we can see the accumulation of signals at lower energies because of the wall effect and the neutron moderation by the surrounding material.

\begin{figure}[!h]
\centering
\includegraphics[width=90mm]{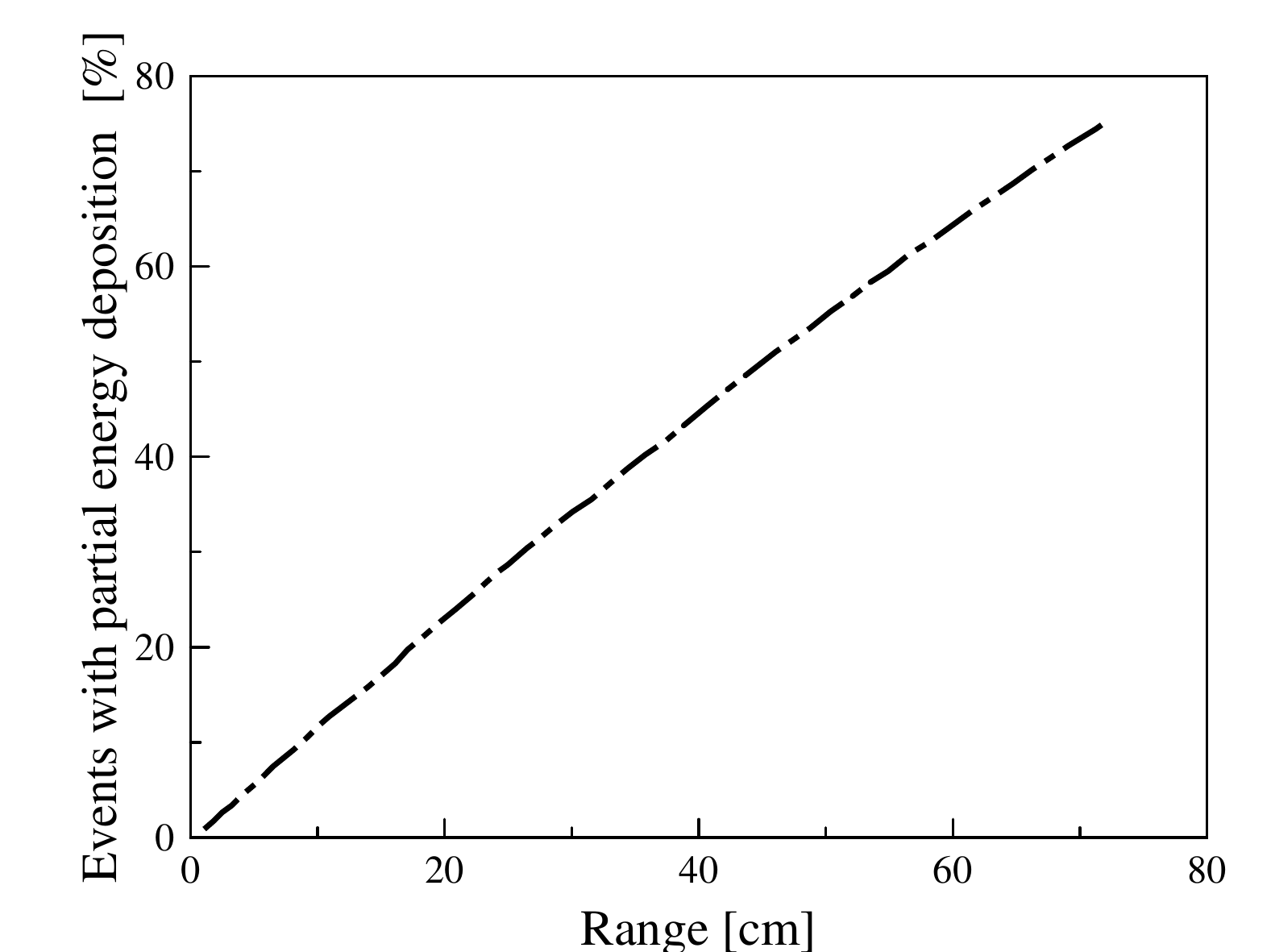}
\caption{Calculated fraction of recorded events ($\%$), with partial energy deposition due to the wall effect as a function of the range of the reaction products, for the SPC with a diameter of 130 cm.}
\label{wall}
\end{figure}

In case the detector being used as a neutron spectrometer and not as a simple neutron counter, the wall effect plays an important role. The wall effect depends on the range of the produced particles and on the size of the detector. 
In figure \ref{wall} we present the wall effect for the geometry of the SPC as a function of the sum of the ranges of the produced charged particles. The calculation was done using a simplified Monte Carlo simulation; we considered a homogeneous distribution of the interaction point inside the spherical volume and an isotropic, back-to-back emission of the reaction products. If any of the two fragments touches the detector shell, the event is accounted as "lost due to wall effect". The ranges depend on the neutron energy and on the gas pressure. 
As an example, for pure $N_2$ gas at $500\ mbar$ and thermal neutrons, the range of the emitted proton is $2.2\ cm$ and the range of the ${}^{14}C^{+}$ is $250\ \mu m$, resulting in $2\%$ of the events lost due to wall effect. The wall effect is more important for the fast neutrons since the range of $1.5\ MeV$ protons is $8.7\ cm$ in pure $N_{2}$ at $500\ mbar$ and the loss of events is about $10\%$.
For neutron energies higher than $2\ MeV$ the ${}^{14}N(n,\alpha){B}^{11}$ reaction is the dominant one for the fast neutron detection. The ranges of the ${}^{11}B$ and the $\alpha$ particles in $500\ mbar$ $N_2$ for $4\ MeV$ neutrons are approximately $0.7\ cm$ and $3\ cm$ respectively and the loss of events is $4\%$.

\FloatBarrier
\section{Comparison with the commercial neutron counters.}
There are many commercially available neutron counters \citep{knoll} mainly in cylindrical shape, with different diameter and length. The tubes are filled with ${}^{3}He$ gas \citep{he3} or ${}^{10}BF_3$ gas \citep{bf3} and the neutrons are detected via the ${}^{3}He(n,p)H^{3}$ and ${}^{10}B(n,\alpha)Li^{7}$ reactions. Both counter types, are very sensitive to thermal neutrons because of the positive Q-values and the high cross sections.

\begin{center}
${}^{3}He(n,p)H^{3}$: $Q=0.76\ MeV$ and $\sigma_{th}=5330\ barn$
\end{center}

\begin{center}
${}^{10}B(n,\alpha)Li^{7}$: $Q=2.79\ MeV$ and $\sigma_{th}=3837\ barn\  (6\%)$
\end{center}

\begin{figure}[!h]
\centering
\includegraphics[width=80mm]{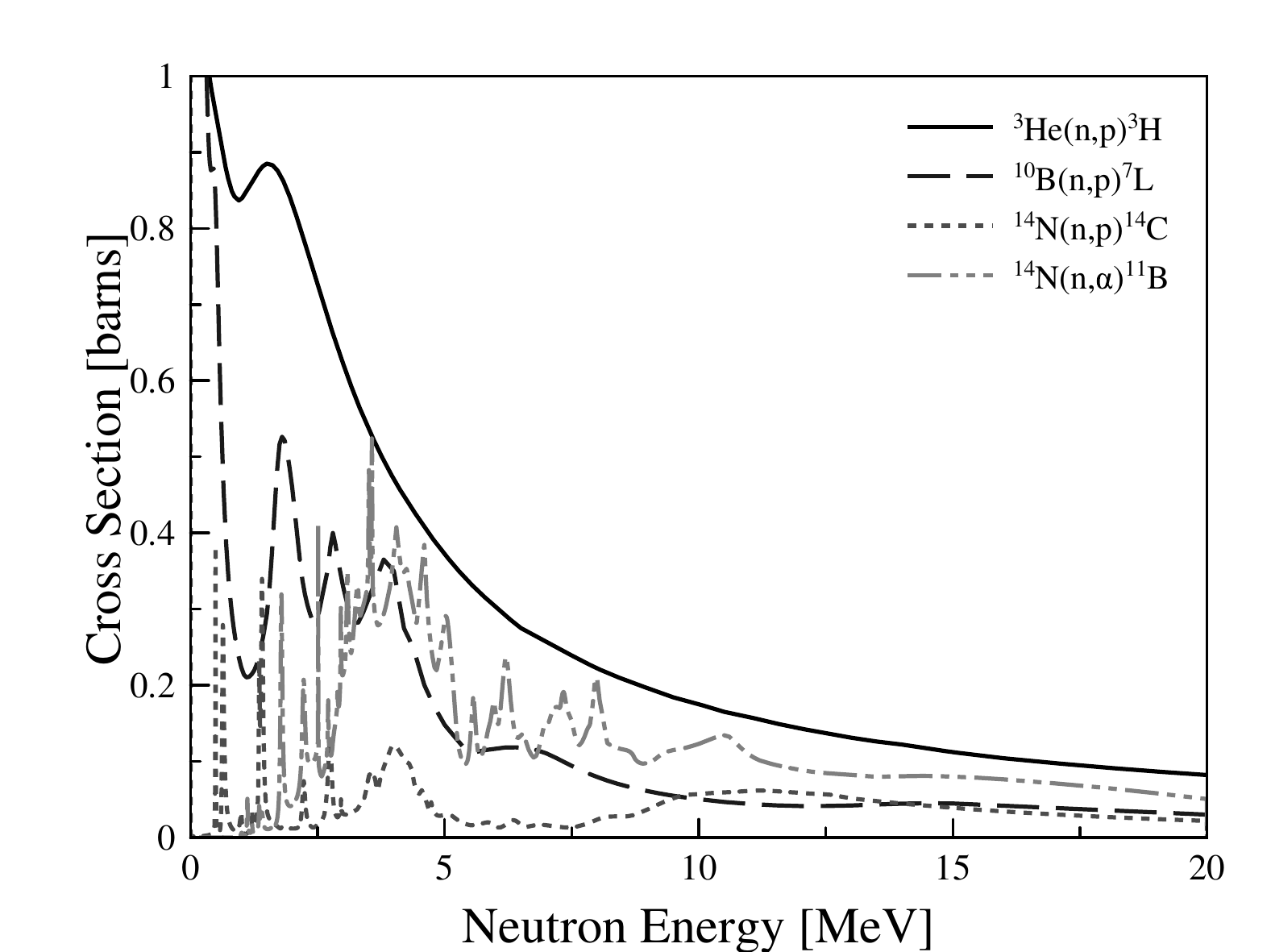}
\caption{ Comparison of the cross sections of the reactions ${}^{3}He(n,p)H^{3}$, ${}^{10}B(n,\alpha)Li^{7}$, ${}^{14}N(n,p){C}^{14}$ and  ${}^{14}N(n,\alpha){B}^{11}$ for fast neutrons of energy up to $20\ MeV$ \citep{endfvII}.}
\label{cs2}
\end{figure}

In comparison with the ${}^{3}He$ and  ${}^{10}BF_3$, the $N_2$ has lower cross section for thermal neutrons, but this disadvantage can be covered sufficiently by the large amount of nitrogen atoms in the spherical counter. Thanks to the large volume of the sphere the wall effect is very small compared with that of the cylindrical tubes. Additionally the cost of the ${}^{3}He$ gas is very high and the ${}^{10}BF_3$ is toxic.
For fast neutrons up to $20\ MeV$ the three isotopes, ${}^{3}He$, ${}^{10}BF_3$ and ${}^{14}N$, have comparable cross sections (figure \ref{cs2}), but the fast neutron events in the cylindrical detectors are masked by the thermal ones. Their detection needs the help of moderators. Whereas the large volume of the spherical detector ($1\ m^{3}$) with the large amount of $N_2$ gas (up to $500\ mbar$) inside, provides the possibility of direct fast neutron spectroscopy. The use of ${}^{3}He$ instead of $N_2$ in the SPC is also suitable for fast neutron spectroscopy; compared to the $N_{2}$ gas it produces nuclear recoils in the same energy region as the $0.76\ MeV$ thermal neutron peak, increasing the background.

\section{Future developments}
The first measurements revealed the capabilities of the spherical detector based on $N_{2}$ as a neutron counter and its potential as a simple neutron spectrometer. The performance of the detector is expected to be improved in the coming tests by considering the following modifications:
\begin{itemize}
\item{operation in higher gas pressure, thus increasing the detection efficiency and minimizing at the same time the wall effect}
\item{reduce the diameter of the electrode ball}
\item{use mixtures of $N_2$ with heavy gases like Ar and Xe. The addition of those gases will reduce the range of the reaction products, also limiting the wall effect}
\end{itemize}

Ongoing simulations show that particle discrimination through pulse shape analysis technique is possible. The separation is based on the fact that the range of protons in the gas is longer than the one of alpha particles of same energy. The radial distance between the start and end point of the particle trajectory (together with the longitudinal diffusion) reflects on the time characteristics of the corresponding pulses. The correlation of the observed rise time and pulse width with the pulse amplitude will allow the discrimination of $(n,p)$ from $(n,\alpha)$ reactions and the rejection of $(n,\gamma)$ reactions, thus simplifying the reconstruction of the neutron spectrum. Furthermore, the wall effect is much less important for the $(n,\alpha)$ reaction, so it is expected to extend the range of the spectroscopy to higher neutron energies.

In order to exploit the spectroscopy capabilities of the spherical detector we are planning a series of measurements with monoenergetic neutron beams in facilities like CEA Cadarache or N.C.S.R. Athens.

\begin{figure}[!h]
\centering
\includegraphics[width=90mm]{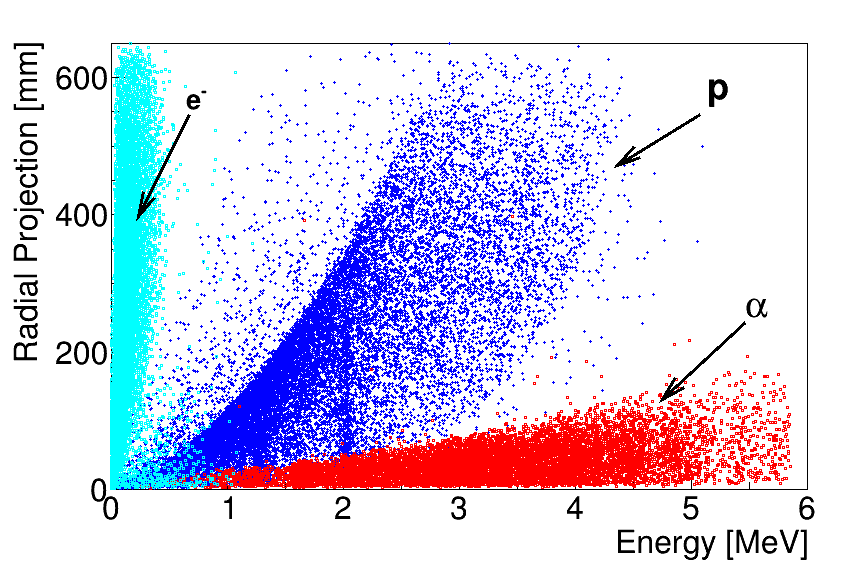}
\caption{Radially projected particle range as a function of the deposited energy, from GEANT4 \citep{agostinelli} simulations for $400\ mbar$ $N_{2}$ gas. The radial range reflects on the pulse characteristics: larger length will give larger rise time and pulse width. The events corresponding to alpha particles, protons or electrons are distributed in three distinguishable populations.}
\label{range}
\end{figure}

\FloatBarrier
\section{Conclusions}
 The SPC can be successfully used as a neutron detector, with pure $N_2$ gas, exploiting the ${}^{14}N(n,p){C}^{14}$ and ${}^{14}N(n,\alpha){B}^{11}$ reactions. We were able to measure the thermal neutron flux from both neutron sources and atmospheric neutrons with success. As for the fast neutrons we were able to measure the ${}^{252}Cf$ and the ${}^{241}Am-{}^{9}Be$ neutrons and to show the fast neutron spectroscopy potential of the spherical detector, for neutron energies up to $20\ MeV$.
 The simple, robust design, the large volume and the possibility of operation at high pressure are important advantages that make the SPC an attractive alternative to the typical cylindrical counters.

\section*{Acknowledgments}
This research/publication has been co-financed by the European Union (European Social Fund – ESF) and Greek national funds through the Operational Program "Education and Lifelong Learning" of the National Strategic Reference Framework (NSRF) - Research Funding Program: “THALIS – HELLENIC OPEN UNIVERSITY- Development and Applications of Novel Instrumentation and Experimental Methods in Astroparticle Physics”.


\bibliography{mybibfile}

\begin{thebibliography}{10}
\expandafter\ifx\csname url\endcsname\relax
  \def\url#1{\texttt{#1}}\fi
\expandafter\ifx\csname urlprefix\endcsname\relax\def\urlprefix{URL }\fi
\expandafter\ifx\csname href\endcsname\relax
  \def\href#1#2{#2} \def\path#1{#1}\fi

\bibitem{giomataris1}
I.~Giomataris, J.~Vergados, Nuclear Instruments and Methods in Physics Research
  Section A 530~(3) (2004) 330 -- 358.
\newblock \href
  {http://dx.doi.org/http://dx.doi.org/10.1016/j.nima.2004.04.223}
  {\path{doi:http://dx.doi.org/10.1016/j.nima.2004.04.223}}.

\bibitem{giomataris2}
{I. Giomataris et al}, Nuclear Physics B - Proceedings Supplements 309~(1)
  (2006) 012031.
\newblock \href
  {http://dx.doi.org/http://dx.doi.org/10.1016/j.nuclphysbps.2005.01.245}
  {\path{doi:http://dx.doi.org/10.1016/j.nuclphysbps.2005.01.245}}.

\bibitem{aune}
{S. Aune et al}, AIP Conf.Proc. 110–8 (2005) 785.

\bibitem{giomataris3}
{I Giomataris et al}, Journal of Instrumentation 3~(09) (2008) P09007.
\newblock \href {http://dx.doi.org/doi:10.1088/1748-0221/3/09/P09007}
  {\path{doi:doi:10.1088/1748-0221/3/09/P09007}}.

\bibitem{andriamonje}
{S. Andriamonje et al}, J. Phys. Conf. Ser. 179 (2009) 012003.
\newblock \href {http://dx.doi.org/10.1088/1742-6596/179/1/012003}
  {\path{doi:10.1088/1742-6596/179/1/012003}}.

\bibitem{savvidis}
{I. Savvidis et al}, {TAUP Conference} 785 (2009) 110–8.

\bibitem{gledenov}
Y.~M. Gledenov, V.~I. Salatski, P.~V. Sedyshev, The 14n(n,p)14c reaction cross
  section for thermal neutrons, Zeitschrift f{\"u}r Physik A Hadrons and Nuclei
  346~(4) (1993) 307--308.
\newblock \href {http://dx.doi.org/10.1007/BF01292522}
  {\path{doi:10.1007/BF01292522}}.

\bibitem{endfvII}
{M.B. Chadwick et al}, {ENDF/B-VII.1} nuclear data for science and technology:
  Cross sections, covariances, fission product yields and decay data, Nuclear
  Data Sheets 112~(12) (2011) 2887 -- 2996, special Issue on ENDF/B-VII.1
  Library.
\newblock \href {http://dx.doi.org/http://dx.doi.org/10.1016/j.nds.2011.11.002}
  {\path{doi:http://dx.doi.org/10.1016/j.nds.2011.11.002}}.

\bibitem{wall1}
R.~Cervellati, A.~Kazimierski, Wall effect in bf3 counters, Nuclear Instruments
  and Methods 60~(2) (1968) 173 -- 178.
\newblock \href
  {http://dx.doi.org/http://dx.doi.org/10.1016/0029-554X(68)90407-2}
  {\path{doi:http://dx.doi.org/10.1016/0029-554X(68)90407-2}}.

\bibitem{wall2}
S.~Shalev, Z.~Fishelson, J.~Cuttler, The wall effect in 3he counters, Nuclear
  Instruments and Methods 71~(3) (1969) 292 -- 296.
\newblock \href
  {http://dx.doi.org/http://dx.doi.org/10.1016/0029-554X(69)90317-6}
  {\path{doi:http://dx.doi.org/10.1016/0029-554X(69)90317-6}}.

\bibitem{iso}
{ISO 8529-1}, International Organization for Standardization, Geneva,
  Switzerland (2001).

\bibitem{knoll}
G.~Knoll, Radiation Detection and Measurement, John Wiley \& Sons, 2010.

\bibitem{he3}
R.~Batchelor, R.~Aves, T.~H.~R. Skyrme, Helium‐3 filled proportional counter
  for neutron spectroscopy, Review of Scientific Instruments 26~(11).

\bibitem{bf3}
I.~L. Fowler, P.~R. Tunnicliffe, Boron trifluoride proportional counters,
  Review of Scientific Instruments 21~(8) (1950) 734--740.
\newblock \href {http://dx.doi.org/http://dx.doi.org/10.1063/1.1745700}
  {\path{doi:http://dx.doi.org/10.1063/1.1745700}}.

\bibitem{agostinelli}
{Agostinelli, S. et al}, Nucl. Instrum. Meth. A506 (2003) 250--303.
\newblock \href {http://dx.doi.org/10.1016/S0168-9002(03)01368-8}
  {\path{doi:10.1016/S0168-9002(03)01368-8}}.

\end{thebibliography}

\end{document}